\documentstyle[aps,aps10,floats]{revtex}

\begin{document}
\title{\bf Phenomenological Study of Strong Decays of Heavy Hadrons in Heavy 
Quark Effective Theory}

\author{N.
Tr\'egour\`es\footnote{Present address: Laboratoire de Physique et 
Mod\'elisation des Milieux Condens\'es
CNRS, Maison des Magist\`eres, Universit\'e Joseph Fourier
B.P. 166, 38042 Grenoble Cedex 9, France.} \\ {\it Department of Physics, Old 
Dominion University, Norfolk, VA 23529}\\
 {\it and Thomas Jefferson National Accelerator Facility, 12000 Jefferson Avenue}\\
  {\it Newport News, VA 23606} \\ and \\ W. Roberts\footnote{On leave from the Department of Physics, Old
Dominion University, Norfolk, VA 23529, and from Thomas Jefferson National 
Accelerator Facility, 12000 Jefferson Avenue, Newport News, VA 23606} \\ {\it National Science Foundation, 4201
Wilson Boulevard, Arlington, VA 22230} }
\maketitle
\thispagestyle{empty}
\begin{abstract}
The application of the tensor formalism of the heavy quark effective theory (HQET) at  
leading order to strong decays of heavy hadrons is presented. Comparisons 
between experimental and theoretical predictions of ratios of decay rates 
for $B$ mesons, $D$ mesons and kaons are given. The 
application of HQET to strange mesons presents 
some encouraging results. The spin-flavor symmetry is used to 
predict some decay rates that have not yet been measured. 

\vspace{5mm}

\flushright{JLAB-THY-99-34}
\end{abstract}


\section{Introduction}

In a recent article \cite{wr}, the formalism of the heavy quark effective 
theory (HQET) \cite{phyrep,Georgi,Smartp1} was applied to the strong decays of heavy hadrons. In that 
article, it was shown that the results for the ratios of decay rates obtained 
by Isgur and Wise \cite{iw1} in their spin counting arguments, and by other 
authors in the combined HQET/chiral perturbation theory \cite{goity,falk1}, were 
reproduced. It was also shown that the treatment of decays beyond the $S$- and 
$P$- wave pion emissions of heavy hadrons were relatively easy to handle.

In this article, we test the formalism by applying it to the measured 
decays of charmed and beauty hadrons. However, since the data in these two 
sectors are limited, we also examine strange hadrons, assuming that we can 
treat the strange quark as a heavy one. This has been done by other authors in 
the past, with reasonable success \cite{mannel}, and we find that our formalism, applied to 
strange hadrons, also works surprisingly well.

In the next section, we briefly review the salient points of the application 
of HQET to the strong decays of hadrons. In section 3 we present our results, 
while in section 4 we give our conclusions. Note that most of the experimental 
results presented in section 3 are obtained from the Particle Data Group 
\cite{PDG}, unless a specific reference is given. For the excited $B$ mesons, 
data are taken from \cite{B-mass,B-mass2,B-mass3,B-mass4}.

\section{Matrix Elements of Strong Decays}

For the sake of completeness, we include here the salient points of HQET as
applied to the strong decays of heavy hadrons. The decay amplitude for
\begin{eqnarray}
H_Q&\longrightarrow& H'_Q+H_l,
\end{eqnarray}
where $H_l$ is a light hadron is given by
\begin{eqnarray}
{\mathcal{M}}&=&<H_lH'_Q|{\mathcal{O}}_s|H_Q>,
\end{eqnarray}
where ${\mathcal{O}}_s$ is the operator responsible for the strong decay. 
Unlike electroweak processes, we do not know the explicit form of 
${\mathcal{O}}_s$. It is expected to be a complicated object involving non-perturbative 
QCD acting on composite, strongly-interacting particles. The only thing we know 
is that ${\mathcal{O}}_s$ is a Lorentz scalar operator as well as an operator that is a  
singlet in all flavors of quarks. Nevertheless we are free to parametrize 
${\mathcal{O}}_s$ in a way that is useful for our purposes. Since HQET helps us to deal with the 
heavy quark field, it is of interest to focus on the heavy quark part of 
${\mathcal{O}}_s$.  Without loss of generality, we parametrize 
${\mathcal{O}}_s$ as \cite{wr}
\begin{eqnarray}
{\mathcal{O}}_s&=&\sum_i{\overline{Q}\Gamma_iQL_i},\label{os}
\end{eqnarray}
where $Q$ is the heavy quark field, $\Gamma_i$ is one of the 16 matrices $I$, $\gamma_{\mu}$, 
$\sigma_{\mu\nu}$, $\gamma_{\mu}\gamma_5$, $\gamma_5$ and $L_i$ contains all 
our ignorance of the dynamics of the light degrees of freedom. $L_i$ has the 
same Lorentz structure as $\Gamma_i$ so that ${\mathcal{O}}_s$ is a Lorentz 
scalar operator. Only heavy quark loop terms are omitted in this 
parametrization of ${\mathcal{O}}_s$, but these contributions are suppressed 
by powers of $1/m_Q$. 

In the heavy quark limit, the heavy quark does not recoil or flip its spin 
during the decay, thus acting as a spectator. The only possible term that 
contributes to the strong decay operator in this case is the term $\Gamma_i=I$ 
\cite{wr}, $i.e.$,
\begin{equation}
{\mathcal{O}}_s\longrightarrow\overline{h}IhL, m_Q\longrightarrow\infty.
\end{equation}
where $L$ is an unknown scalar operator that acts on the brown muck component of the hadron, 
and $h$ is the effective heavy 
quark field \cite{phyrep,Georgi,Smartp1}. In the heavy quark limit the decay 
amplitude thus becomes 
\begin{eqnarray}
{\mathcal{M}}&=&<H_lH'_Q|\overline{h}hL|H_Q>.
\end{eqnarray}

In order to calculate ${\mathcal{M}}$, we use the representations of states
developed by Falk \cite{Falk:representation}.
For concreteness let us examine the example of a meson decay. Consider the 
decay of the meson doublet $(1^+,2^+)$, $J^P_\ell={\frac{3}{2}}^+$, to the meson 
doublet $(0^-,1^-)$, $J^P_\ell={\frac{1}{2}}^-$ with the emission of a single pion.
The four possible decays are 
\begin{eqnarray}
D_1 &\rightarrow & D +\pi,\nonumber\\ 
D_1 &\rightarrow & D^* +\pi,\nonumber\\ 
D_2 &\rightarrow & D +\pi,\nonumber\\ 
D_2 &\rightarrow & D^* +\pi.\nonumber 
\end{eqnarray} 
At leading order in HQET ($i. e.$, in the limit that the mass of the heavy quark goes 
to infinity) the matrix elements are \cite{wr}
\begin{eqnarray}
{\mathcal{M}}_{1D} &=&\label{me1} <D\pi (p) | \bar c c L| D_1> =  \sqrt{m_{D} m_{D_1}} 
Tr[\gamma_5 T_{\mu}\bar {\cal D}{\cal D}^{\mu}_1], \\ 
{\mathcal{M}}_{1D^*} &=& \label{me2}<D^*\pi (p) | \bar c c L| D_1> =  \sqrt{m_{D^{*}} 
m_{D_1}} Tr[\gamma_5 T_{\mu}\bar {\cal D}^{*} {\cal D}^{\mu}_1], \\ 
{\mathcal{M}}_{2D} &=& \label{me3}<D\pi (p) | \bar c c L|D_2> =  \sqrt{m_D m_{D_2}} 
Tr[\gamma_5 T_{\mu} \bar {\cal D}{\cal D}^\mu_2],\\ 
{\mathcal{M}}_{2D^*} &=&\label{me31} <D^*\pi (p) | \bar c cL |D_2> =  \sqrt{m_{D^{*}} 
m_{D_2}}
 Tr[\gamma_5  T_{\mu} \bar {\cal D}^{*} {\cal D}^\mu_2],
\end{eqnarray}
where ${\cal D}$, ${\cal D}^*$, ${\cal D}^{\mu}_1$ and ${\cal D}^\mu_2$ are the 
tensor representations of $D$, $D^*$, $D_1$ and $D_2$, respectively, \cite{Falk:representation}.

The only non-redundant form allowable for 
$T_{\mu}$ is \cite{wr}
\begin{eqnarray}
T_{\mu}&=&\alpha p_{\mu}\not\!p,
\end{eqnarray}
where $p$ is the four momentum of the pion. The constant $\alpha$ parametrizes our ignorance of the
non-perturbative aspects of these decays. It is a constant that may be estimated in a quark model, for instance,
but it is one about which HQET can say nothing. The form written above is valid for all four decays
considered. Consequently, ratios of decay rates of any of these processes are fully 
determined by the heavy quark formalism. Moreover, due to the flavor symmetry and our choice of normalization, the coupling 
constant $\alpha$ would be the same in the case of the $b$-flavored as well as $s$-flavored 
mesons (if the $s$ quark is treated as heavy). This means that, had we used  $B, B^*, 
B_1, B_2$ or $K, K^*, K_1, K_2$ instead of $D, D^*, D_1, D_2$ the expressions 
(\ref{me1}) to (\ref{me31}) would have been exactly the same but for the mass factors.

\section{Results} \label{sec2}

\subsection{$D_1$ and $D_2$ }

We first apply HQET to $D$ mesons. Using the heavy quark limit is not as reasonable as 
it would be for $B$ mesons, but more data are available. The experimental 
status of $D$ mesons is summarized in table~\ref{mesonsd}. $D_1$ and $D_2$ are 
believed to correspond to the doublet $(1^+,2^+)$, $J^P_\ell={\frac{3}{2}}^+$ while $D$ and 
$D^*$ correspond to the doublet $(0^-,1^-)$, $J^P_\ell={\frac{1}{2}}^-$.

\squeezetable

\begin{table}
\begin{center}
\begin{tabular}{|c|c|c|c|c|c|c|} \hline 
 $J^P_\ell$ & $J^P$ & State & Mass (MeV)& $\Gamma$(MeV)&$\Gamma_{D\pi}$(MeV)&$\Gamma_{D^*\pi}$(MeV)   \\ \hline 
 ${\frac{1}{2}}^-$ & $0^-$ & $D^0$ &$ 1864.6\pm0.5$  & $\tau = 0.415\pm0.004 ps$&$-$&$-$\\ 
 		 & 	 & $D^\pm $ & $1869.3\pm0.5$& $\tau = 1.057\pm0.015 ps$&$-$&$-$ \\ \hline
 		 & $1^-$ & $D^{0*}$& $2006.7\pm0.5$& $<2.1$&$<1.3$&$-$ \\ 
 		 & 	 & $D^{0\pm}$ &$2010.0\pm0.5$& $<0.131$&$<0.130$&$-$ \\ \hline\hline
 ${\frac{3}{2}}^+$ & $1^+$ & $D_1(2420)^0$ &$2422.2 \pm 1.8$ & $18.9^{+4.6}_{-3.5}$&not seen&seen \\ 
 		 & 	 & $D_1(2420)^\pm$ & $2427 \pm 5$ & $28 \pm 8$ &not seen&seen\\ \hline
 		 & $2^+$ & $D_2(2460)^0$ & $2458.9 \pm 2.0$ & $23 \pm 5$ & seen&seen\\ 
 		 & 	 & $D_2(2460)^\pm$ & $2459 \pm 4$ & $25^{+8}_{-7}$ &seen&seen\\ \hline
\end{tabular}
\caption{Summary of $D$ meson experimental status. $J^P_\ell$ is the spin-parity of the light 
degrees of freedom according to the HQET.}\label{mesonsd}
\end{center}
\end{table}

$D_1$ and $D_2$ are constrained by spin-flavor symmetry to decay via $D$-wave pion 
emission. The channels which are allowed are
\begin{eqnarray}
D_2 & \rightarrow & D^* + \pi, \\
D_2 & \rightarrow & D + \pi, \\
D_1 & \rightarrow & D^* + \pi. 
\end{eqnarray}
For these decays, we obtain the well known 
results 
of table~\ref{t1-Dmeson} \cite{D-ratios}.
\begin{table}
\begin{center}
\begin{tabular}{|c|c|c|} \hline 
Ratios of widths & Experiment  &  Heavy quark prediction  \\ \hline 
\(\frac{\Gamma( D_2^0 \rightarrow D^\pm +\pi^\mp)}{\Gamma(D_2^0 \rightarrow 
D^{*\pm} +\pi^\mp)}\) & \(2.3 \pm 0.6\) &2.3 \\ \hline
\(\frac{\Gamma( D_2^\pm \rightarrow D^0 +\pi^\pm)}{\Gamma(D_2^\pm \rightarrow 
D^{*\pm} +\pi^\mp)}\) & \(1.9 \pm
1.1 \pm 0.3\) &2.3 \\ \hline
 \(\frac{\Gamma( D_1^{0})}{\Gamma( D_2^0)}\) & \(0.82 \pm 0.5\) & 0.32 \\ \hline
 \(\frac{\Gamma( D_1^{\pm})}{\Gamma( D_2^\pm)}\) & \(1.12 \pm 0.5\) & 0.35 \\ \hline
\end{tabular}
\caption{Ratios of partial and total widths for $D_1$ and $D_2$ mesons. The numbers 
in the second column are the experimental ratios, while the numbers in the third column 
are the leading order HQET predictions.}\label{t1-Dmeson}
\end{center}
\end{table}

Clearly, ratios of partial widths work very well, while ratios of total widths, which 
are just the sum of partial widths, only agree at the level of one to two
standard deviations.
Different interpretations for this discrepancy have been given in the literature. One 
possibility is to assume that the widths of $D_1$ could receive a contribution from two
 pion decay (non-resonant or through an intermediate $\rho$ meson) to the ground state 
 $D$ \cite{2pi}
\begin{eqnarray}
D_1 & \rightarrow & D + \pi+ \pi,\\
D_1 & \rightarrow & D + \rho \rightarrow  D + \pi+ \pi.
\end{eqnarray}
This could broaden the $D_1$ if there is no analogous enhancement in the $D_2$ decay 
and consequently increase the ratio $\frac{\Gamma(D_1)}{\Gamma(D_2)}$ without 
changing partial ratios. However, up to now, there is no experimental evidence of such an 
effect. Another option is to assume a mixing of the narrow $D_1$ with the broad  $D'_1$ 
($J^P_\ell={\frac{1}{2}}^+$) \cite{D-ratios}. Since $D'_1$ decays via $S$-wave pion 
emission rather than $D$-wave, it is expected to be much broader. Mixing of 
states is 
forbidden by the spin symmetry but is allowed when $1/m_Q$ effects are included. 
However, so far, there is no evidence of an $S$-wave 
component in the decay of $D_1 \rightarrow D^* + \pi$ \cite{Swave}. 
Another explanation, fully consistent with the HQET, has been given by Falk and Mehen 
\cite{falk1}. They argue that the discrepancy of total width ratios could be due 
to terms of subleading order in the $1/m_Q$ expansion.  They found that the experimental 
ratio $\frac{\Gamma(D_1)}{\Gamma(D_2)}$ can be predicted without involving mixing of 
states, (even though they studied the possibility of such mixing). These corrections from 
subleading order in the heavy quark expansion are beyond the scope of 
this manuscript and are left for a possible future work.

\subsection{$B_1$ and $B_2$}\label{mB}

Because of the flavor-symmetry, ratios of decay rates calculated for $D_1$ and $D_2$ 
are also valid for 
$B_1$ and $B_2$ as long as we use the correct masses and pion momenta. Moreover, 
since $\Lambda_{QCD}/m_b \simeq 0.1$, $1/m_b$ corrections to the leading order should 
be, in this case, much smaller than for the $D$ mesons ($\Lambda_{QCD}/m_c \simeq 0.3$). 

Unfortunately, $B$ mesons are not very well known experimentally. A short 
summary of the experimental status of bottom mesons is given in table~\ref{mesonsb}.
The masses of the $B_1$ and $B_2$ are not precisely known, and only a few decay 
rates are available in the literature. Consequently, since our predictions of widths are 
extremely sensitive to the masses of $B_1$ and $B_2$, we present different ratios using 
different masses found in the literature.

\begin{table}
\begin{center}
\begin{tabular}{|c|c|c|c|c|c|c|} \hline 
 $J^P_\ell$ & $J^P$ & State & Mass (MeV)& $\Gamma$(MeV)& $\Gamma_{B\pi}$(MeV)& $\Gamma_{B^*\pi}$(MeV)    \\ \hline 
 ${\frac{1}{2}}^+$ & $0^-$ & $B^-$ & $5279.2 \pm 1.8$  & $\tau = 1.56\pm0.04$ ps &$-$&$-$\\ 
 		 & 	 & $B^\pm $ & $5278.9 \pm 1.5$ & $\tau = 1.65\pm0.04$ ps &$-$&$-$ \\ \hline
 		 & $1^-$ & $B^{*}$& $5324.8 \pm 1.8$ & $<6$ &no data&$-$\\ \hline\hline
 ${\frac{3}{2}}^+$ & $1^+$ & $B_1$ & $\sim 5700$  & $20\pm?$ &no data& no data \\ \hline
 		 & $2^+$ & $B_2$ & $\sim 5700$ & $25\pm?$ &no data&no data\\ \hline\hline
 ? & ? & $B_j$ &$5698\pm12$  & $128\pm18$  &seen& seen \\ \hline
\end{tabular}
\caption{Summary of $B$ meson  experimental status. $J^P_\ell$ is the spin-parity of the light degrees of freedom according to the HQET.  }\label{mesonsb}
\end{center}
\end{table}

Ratios of partial and total widths for $B_1$ and $B_2$ are given in 
table~\ref{t1-Bmeson}. The first two columns are HQET predictions using 
experimental masses from  \cite{B-mass,B-mass2,B-mass3,B-mass4}. The third 
column contains HQET predictions using masses obtained by applying the spin-flavor symmetry of
HQET. 
This symmetry relates $m_{B_1}$ and $m_{B_2}$ to $m_{D_1}$ and $m_{D_2}$, 
which are precisely known \cite{falk1}. More precisely, this symmetry relates 
$m_{B_2}-m_{B_1}$ to $m_{D_2}-m_{D_1}$, as well as $m_{D_1}-m_D$ to $m_{B_1}-m_B$, modulo
$1/m_Q$ corrections. One therefore expects that the theoretical estimates of the masses should
be accurate up to $1/m_Q$ corrections. Finally the last column gives 
experimental results. At present, there are no experimental errors for the 
widths, thus the only experimentally determined ratio must be interpreted with some caution. 

\begin{table}
\begin{center}
\begin{tabular}{|c|c|c|c|c|}\hline 
 Ratios of  widths     & $m_{B_1}=m_{B_2}=5.68$   &  $m_{B_1}=5.725,m_{B_2}=5.737$  & 
 $m_{B_1}=5.780,m_{B_2}=5.794$ & Experiment\\ \hline 
$\frac{\Gamma( B_2 \rightarrow B +\pi)}{\Gamma(B_2 \rightarrow B^* +\pi)}$ & 1.30&1.17&1.08 & no data\\ \hline
$\frac{\Gamma( B_2 \rightarrow B^* +\pi)}{\Gamma(B_1 \rightarrow B^* +\pi)}$ & 0.60&0.71&0.70  & no data\\ \hline
$\frac{\Gamma( B_2 \rightarrow B +\pi)}{\Gamma(B_1 \rightarrow B^* +\pi)}$ & 0.78&0.82&0.75  & no data\\ \hline
$\frac{\Gamma( B_2)}{\Gamma( B_1)}$ &1.38&1.53&1.45  & $1.25\pm?$\\ \hline
\end{tabular}
\caption{Ratios of partial and total widths for $B_1$ and $B_2$. The numbers in the columns two
and three are the leading order predictions of HQET using experimental masses found in the 
literature while the numbers in column four are predictions using masses obtained via 
relations between $B$ mesons and $D$ mesons.  The numbers in the last column are the 
experimental values.} \label{t1-Bmeson}
\end{center}
\end{table}

From our $D$ meson results, one might expect more reliable ratios of partial 
widths than total widths. In addition, since $\Lambda_{QCD}/m_b \simeq 0.1$, 
$1/m_b$ corrections should be much smaller than for the $D$ mesons, and 
predictions more reliable. However, due to the 
uncertainties of the masses we cannot really take advantage of the expected 
better convergence of the $1/m_Q$ expansion. As a result we can not make any 
clear prediction, as we note that our results are extremely mass dependent 
but are in reasonable agreement with the only experimental data available.

We can also attempt to predict absolute partial as 
well as absolute total widths for $B_1$ and $B_2$ using $D$ mesons decay 
rates. Let us denote the pion momenta in the processes  $B_2\rightarrow 
B^*\pi$, $B_2\rightarrow B\pi$ and $B_1\rightarrow B^*\pi$ as $\vec p_{2B^*}$,
$\vec p_{2B}$ and $\vec p_{1B}$, respectively (with a similar notation for the
corresponding $D$ decays).
Then the partial widths of $B_1$ and $B_2$ are related to those of the $D_2$ by
\begin{eqnarray}
\frac{\Gamma( B_2 \rightarrow B^*+\pi)}{\Gamma(D_2)}&=& \label{bd1}
\frac{m_{D_2}}{m_{B_2}}\frac{3/5{|\vec p_{2B^*}|}^5 m_{B^*}}{6/15{|\vec p_{2D}|}^5
m_{D}+3/5{|{\vec p_{2D^*}}|^5
m_{D^*}}}, \\
\frac{\Gamma( B_2 \rightarrow B +\pi)}{\Gamma(D_2)}&=& \label{bd2}
\frac{m_{D_2}}{m_{B_2}}\frac{6/15{|\vec p_{2B}|}^5 m_{B}}{6/15{|\vec p_{2D}|}^5
m_{D}+3/5{|{\vec p_{2D^*}}|^5
m_{D^*}}}, \\
\frac{\Gamma( B_1 \rightarrow B^* +\pi)}{\Gamma(D_2)}&=& \label{bd3}
\frac{m_{D_2}}{m_{B_1}}\frac{{|\vec p_{1B}|}^5 m_{B^*}}{6/15{|\vec p_{2D}|}^5
m_{D}+3/5{|{\vec p_{2D^*}}|^5 m_{D^*}}}.
\end{eqnarray}

It would have been simpler to simply use the equations for the ratios of partial widths, 
but the branching ratios for $D_2\rightarrow D^{(*)}\pi$ and $D_1\rightarrow 
D^*\pi$ have not yet been determined. 
We remind the reader that equations (\ref{bd1}), (\ref{bd2}) and (\ref{bd3}) 
are strictly leading order predictions of HQET. We have not included any 
$1/m_Q$ effects. This means that $B_1$ and $B_2$ as well as $D_1$ and $D_2$ decay only via $D$-wave pion emission.

Decay rates of $B_1$ and $B_2$ predicted by equations (\ref{bd1}), (\ref{bd2}) and 
(\ref{bd3}) are shown in table~\ref{t2-Bmeson}. The theoretical errors are obtained 
using experimental uncertainties for $D$ meson widths. Here again, the  first two 
columns are leading order predictions of HQET using experimental masses while the third 
column shows predictions using masses obtained via relations between $B$ mesons and $D$ 
mesons. The last column contains experimental results.

\begin{table}
\begin{center}
\begin{tabular}{|c|c|c|c|c|} \hline 
 Widths (MeV)     & $m_{B_1}=m_{B_2}=5.68$   &  $m_{B_1}=5.725,m_{B_2}=5.737$  & 
 $m_{B_1}=5.780, m_{B_2}=5.794$ & Experiment\\ \hline 
\(\Gamma( B_2 \rightarrow B +\pi)\)  &$4\pm 1$ &$8\pm 2$& $15\pm 5 $&no data\\ \hline
\(\Gamma(B_2 \rightarrow B^* +\pi)\) &$3\pm 1$ &$7\pm 2$&$14\pm 4$ &no data\\ \hline
\(\Gamma( B_2)\)& $7\pm 2$ &$15 \pm 4$ &$29 \pm 9$& $25\pm?$\\ \hline
\(\Gamma( B_1 \rightarrow B +\pi) \) &0 &0 &0 & no data \\ \hline
\(\Gamma( B_1 \rightarrow B^* +\pi) \) &$5 \pm 2$ & $ 10 \pm 3$ & $ 20 \pm 6$ &no data\\ \hline
\(\Gamma( B_1 )\) & $5 \pm 2$ & $ 10 \pm 3$ & $ 20 \pm 6$ &$20\pm?$\\ \hline
\end{tabular}
\caption{Widths of $B_1$ and $B_2$ mesons. The numbers in the two first columns contain 
leading order predictions of HQET using experimental masses while the numbers in the third 
column present predictions using masses obtained via relations between $B$ mesons and $D$ 
mesons. The last column shows experimental data.}\label{t2-Bmeson}
\end{center}
\end{table}

The decay rates of the two first columns (the ones using experimental masses) are 
somewhat lower than the experimental values. On the other hand the third column, the 
one using the spin-flavor symmetry for the masses, is in good agreement with the 
experimental results. At this stage we do not make any definite conclusions since the 
experimental data are not sufficiently precise. However it will be interesting to 
follow the evolution of the experimental masses of $B_1$ and $B_2$ to see if they get 
closer to what we expect from the heavy quark symmetry or not.

\subsection{Strange mesons}

Using the heavy quark limit for $B$ and $D$ mesons is a reasonable approximation, but 
one might also ask how well can the strange quark be described in this limit. 
In order to partially answer that question, we apply the  heavy quark limit to kaons. 
Doing this allows us to test the heavy quark predictions in more excited states 
than previously.
Nevertheless one has to keep in mind that, in the following results, we consider the 
$s$ quark as heavy, which is far less reasonable than for the $b$ or $c$ quark. This has
been done with some success by Mannel \cite{mannel}, for example.

In order to apply HQET to strange mesons we first need to identify  heavy meson 
doublets for excited kaons. We do this in the following manner. First we retain states 
with data available for pion emission, namely $K^{**} \rightarrow K^{(*)}+\pi$. Then, 
we search for two states which have a good set of spin-parity (~for example $(0^-,1^-)$ 
or  $(1^+,2^+)$~) with about $100-150$ MeV mass splitting. This splitting,  which 
arises from subleading order in heavy quark expansions, is expected to be not more than 
$200$ MeV, as quark models indicate \cite{quarkmodel}. Once such a pair of states is 
found, we identify them as possible members of a doublet. We then compare their 
experimental and theoretical ratios of decay rates for pion emission. If they are 
similar, then perhaps we have made a believable identification of a doublet.

\subsubsection{ Positive parity kaons.}\label{ppk}

The status of positive parity kaon resonances is summarized in table~\ref{mesonskplus}. 
From these nine resonances we found four states as possible members of two doublets.

\begin{table}
\begin{center}
\begin{tabular}{|c|c|c|c|c|c|} \hline 
  $J^P$ & State & Mass (MeV)& $\Gamma$(MeV) & $\Gamma_{K\pi}$(MeV)& $\Gamma_{K^*\pi}$(MeV)   \\ \hline 
 $1^+$ & $K_1(1270)$ & $1273\pm 7$ & $90 \pm 20$  & no data &$14.5\pm 5.5$ \\ 
 $1^+$ & $K_1(1400)$ & $1402\pm 7$ & $195 \pm 25$  & no data &$183\pm 26$ \\ 
 $0^+$ & $K_0(1430)$ & $1429\pm 9$ & $287 \pm 10 \pm 21$  & $270\pm40$ & no data \\ 
 $2^+$ & $K^*_2(1430)$ & $1425.6\pm 1.3$ & $98.5\pm2.7$  &  $52\pm 5$ & $26.5\pm3$\\\hline
 $1^+$ & $K_1(1650)$ & $1650\pm 50$ & $150 \pm 50$  & no data& no data \\ 
 $0^+$ & $K^*_0(1950)$ & $1945\pm 30$ & $201 \pm 34 \pm79$  & $104\pm71$ & no data\\ 
 $2^+$ & $K^*_2(1980)$ & $1975\pm 33$ & $373 \pm 33\pm60 $  & seen & no data \\\hline
 $4^+$ & $K^*_4(2045)$ & $2045\pm 9$ & $198 \pm 30$  & $20 \pm 4$ & no data\\ 
 $3^+$ & $K_3(2320)$ & $2324\pm 24$ & $150 \pm 30$  & no data  & no data\\\hline
 \end{tabular}
\caption{Experimental status of positive parity kaon resonances.}\label{mesonskplus}
\end{center}
\end{table}

The pair $(K_1(1270), K_2(1430))$ could be the lowest lying
  $(1^+,2^+)$ doublet ($J^P_\ell={\frac{3}{2}}^+$). The comparison between experimental 
  ratios of widths and the HQET predictions are shown in table~\ref{t1-kaons}. The 
  second column of this table shows the experimental ratios while the third column 
  shows the HQET predictions. Experimental uncertainties on these ratios are of the order of 
  100\% except for the last ratio. The HQET 
  predictions are in good agreement with experimental ratios for the first two cases. 
  On the last ratio, however, our prediction is too high by a factor of two. In this 
  specific case it would be particularly interesting to calculate $1/m_Q$ corrections 
  (including mixing effects) and see if they can decrease the ratio to something closer 
  to the experimental one.

The pair $(K_0^*(1430), K_1(1400))$ could be the lowest lying $(0^+,1^+)$ 
doublet 
($J^P_\ell={\frac{1}{2}}^+$). The comparison between experimental ratios of widths and the 
HQET predictions are shown in table~\ref{t1-kaons}.  In this case, the experimental 
uncertainty on the ratio of rates is smaller than the previous pair of states. However 
the width of $K_1(1400)$ is not very well known ($i.e.$ different experiments give 
quite different widths). We chose to use the width given by Daum {\it et al.} \cite{Daum} 
( $\Gamma (K_1(1400))=195\pm 25$ MeV) and not the widths from \cite{PDG} because the 
branching ratio is extracted from this experiment. This gives us a partial width of
$\Gamma (K_1(1400) \rightarrow K^* + \pi)=183\pm 26$ MeV. The HQET prediction is lower 
than the experimental ratio. However, since one expects $1/m_Q$ corrections to be 
important in the case of the $s$ quark, we believe that the previous identification is 
reasonable. On the other hand, the identification is based on only one ratio. Therefore 
in this particular case more data are crucial in order to validate the doublet.

\begin{table}
\begin{center}
\begin{tabular}{|c|c|c|} \hline 
 Ratio of widths     & Experiment&  Heavy quark prediction   \\ \hline 
$\frac{\Gamma(K_1(1270)  \rightarrow K^* +\pi)}{\Gamma( K_2(1430) \rightarrow K^* +\pi)} $ &
$0.55 \pm 0.45$&0.36\\ \hline
$\frac{\Gamma(K_1(1270)  \rightarrow K^* +\pi)}{\Gamma( K_2(1430) \rightarrow K +\pi)} $ &
$0.3 \pm 0.4$&0.35\\ \hline
$\frac{\Gamma(K_2(1430)  \rightarrow K^* +\pi)}{\Gamma( K_2(1430) \rightarrow K +\pi)} $ &
$0.51 \pm 0.04$&0.99\\ \hline\hline
$\frac{\Gamma(K_0(1430)  \rightarrow K +\pi)}{\Gamma( K_1(1400) \rightarrow K^* +\pi)} $ &
$1.45 \pm 0.50$  & 0.80\\ \hline
\end{tabular}
\caption{Ratios of widths for the $(1^+,2^+)$  
doublet $(K_1(1270),K_2(1430))$, and for the $(0^+,1^+)$ doublet $(K_0(1430),K_1(1400))$.
The second column shows the experimental data while the third column shows the leading
order HQET predictions.}\label{t1-kaons}
\end{center}
\end{table}

We comment here on the choice of the $K_1(1270)$ as the member of this multiplet,
instead of the $K_1(1400)$. The HQET prediction is that the two members of a multiplet
will have the same total width, if they are degenerate. Breaking this degeneracy would yield
slightly different widths. Since the $K_2$ has a total width of 98.5 MeV, and the state at
1270 has a total width of 90 MeV, we thought that this was a closer match than a total width
of 195 MeV. In addition, the widths for the pion-emission decays of these states also give
some clue as to how they should be assigned. 

A better test would be the partial waves in the pion decays of these states, as
the members of the $(1^+, 2^+)$ doublet should decay only through $D$-wave pion emission. In
the absence of such information, the only criterion is the total width, which leads us to pair
the states as we have. In addition, we can examine the comparison between the HQET prediction 
and experiment if we switch the assignments of the two $1^+$ states. This comparison is shown in 
table \ref{switch}. As can be seen, the ratios obtained with this switch are in complete disagreement with the
HQET predictions. Finally, we note Isgur \cite{Isgur} has placed these states in the same doublets that we
have.

\begin{table}
\begin{center}
\begin{tabular}{|c|c|c|} \hline 
 Ratio of widths     & Experiment&  Heavy quark prediction   \\ \hline 
$\frac{\Gamma(K_1(1400)  \rightarrow K^* +\pi)}{\Gamma( K_2(1430) \rightarrow K^* +\pi)} $ &
$6.9\pm 0.8$&1.40\\ \hline
$\frac{\Gamma(K_1(1400)  \rightarrow K^* +\pi)}{\Gamma( K_2(1430) \rightarrow K +\pi)} $ &
$3.5\pm 0.4 $&1.38\\ \hline
$\frac{\Gamma(K_2(1430)  \rightarrow K^* +\pi)}{\Gamma( K_2(1430) \rightarrow K +\pi)} $ &
$0.51 \pm 0.04$&0.99\\ \hline\hline
$\frac{\Gamma(K_0^*(1430)  \rightarrow K +\pi)}{\Gamma( K_1(1270) \rightarrow K^* +\pi)} $ &
$18.7\pm 7.0$  & 1.45\\ \hline
\end{tabular}
\caption{Ratios of widths for the $(1^+,2^+)$  
doublet $(K_1(1400),K_2(1430))$, and for the $(0^+,1^+)$ doublet $(K_0(1430),K_1(1270))$.
The second column shows the experimental data while the third column shows the leading
order HQET predictions. The assignments of the two $1^+$ states are switched from
table \ref{t1-kaons}.}\label{switch}
\end{center}
\end{table}

It may appear discouraging that only two possible doublets were found since nine 
positive parity kaons are experimentally known. The reason is that only six states have 
experimentally measured pion emission decay rates, and only four states out of nine 
could be tested as members of a doublet. We believe that it is quite encouraging to be 
able to find two possible doublets ($i.e.$ four states) with only four states tested.

\subsubsection{Negative parity kaons}

The experimental status of the known negative parity kaon resonances is summarized in 
table~\ref{mesonskmoins}. Including $K$ and $K^*$, thirteen negative parity resonances 
are experimentally known. Unfortunately, data for pion emission are available for only four of
these states. Consequently, we are unable to clearly identify any doublets. However, we are
still able to ratios of decay widths for two states. However in each case, the multiplet 
partner has to be found.

\begin{table}
\begin{center}
\begin{tabular}{|c|c|c|c|c|c|} \hline 
 $J^P$ & State & Mass (MeV)& $\Gamma$(MeV) & $\Gamma_{K\pi}$(MeV)& $\Gamma_{K^*\pi}$(MeV) \\ \hline 
 $0^-$ & $K^{\pm}$ & $493.677\pm0.013$ & $-$  & $-$ & $-$\\ 
 $0^-$ & $K^{0}$ & $497.672\pm0.031$ & $-$  & $-$ & $-$\\
 $1^-$ & $K^{*\pm}(892)$ & $891.66\pm0.26$ & $50.8\pm0.9$  & $~100$\% & $-$\\ 
 $1^-$ & $K^{*0}(892)$ & $896.1\pm0.28$ & $50.5\pm0.6$  & $~100$\%& $-$ \\\hline
 $1^-$ & $K^*(1410)$ & $1414\pm 15$ & $232 \pm 21$  & $15\pm3$ & $>93\pm9$ \\ 
 $0^-$ & $K(1460)$ & $\sim 1460$ & $\sim 260$  &  no data & $\sim 109$ \\ 
 $2^-$ & $K_2(1580)$ & $\sim 1580$ & $\sim 110$  & no data &  seen \\ 
 $1^-$ & $K^*(1680)$ & $1717\pm 27$ & $322\pm 110$  &  $124\pm43$ & $96\pm35$\\
 $2^-$ & $K_2(1770)$ & $1773\pm 8$ & $186 \pm 14$  & no data &  seen \\ 
 $3^-$ & $K^*_3(1780)$ & $1776\pm 7$ & $159 \pm 21$  & $29\pm4$ & $32\pm9$ \\ 
 $2^-$ & $K_2(1820)$ & $1816\pm 13$ & $276 \pm 35 $  & no data & seen  \\
 $0^-$ & $K(1830)$ & $1830\pm 3$ & $250$  & no data  & no data\\\hline
 $2^-$ & $K_2(2250)$ & $2247\pm 17$ & $180 \pm 30 $  & no data & no data\\ 
 $5^-$ & $K^*_5(2380)$ & $2382\pm 14 \pm 19$ & $178 \pm 37 \pm 32$  & $11\pm5$ & no data\\ 
 $4^-$ & $K_4(2500)$ & $2490\pm 20$ & $250 $  & no data & no data\\\hline
\end{tabular}
\caption{Experimental status of negative parity kaon resonances.}\label{mesonskmoins}
\end{center}
\end{table}

$K^*(1680)$ could be the $1^-$ state of a $(1^-,2^-)$ doublet $(J^P_\ell={\frac{3}{2}}^-)$. 
The comparison between experimental ratios of widths
and the HQET predictions can be found in table~\ref{t3-kaons}. The second column shows the
experimental ratio, while the third column shows the HQET prediction, which is 
in very good agreement with the experimental ratio. If, instead, we identify this state as the 
$1^-$ state of a $(0^-,1^-)$ doublet, with  
$(J^P_\ell={\frac{1}{2}}^-)$, the HQET prediction for the ratio of partial 
widths is 0.20. This is far from the 
experimental value. In these two scenarios, the possible multiplet partners are the
$K(1460)$ or the $K_2(1580)$. In either case, the partial width for pion emission has not been
measured.

$K_3^*(1780)$ could be identified with the $3^-$ state of a $(2^-,3^-)$ doublet
 $(J^P_\ell={\frac{5}{2}}^-)$. The comparison between experimental ratios of widths and 
 the HQET predictions are also shown in table~\ref{t3-kaons}.
 Here again, the HQET prediction is very close to the experimental ratio. 
 
 For comparison, if we take this state as the $3^-$ state of a $(3^-,4^-)$ doublet 
 $(J^P_\ell={\frac{7}{2}}^-)$ instead of a $(2^-,3^-)$ doublet $(J^P_\ell={\frac{5}{2}}^-)$, 
 the HQET prediction for the ratio of partial widths is 0.5, which is more than two 
 times smaller than the experimental value.
 
There are two possible partners for $K_3^*(1780)$, namely $K_2(1770)$ and 
$K_2(1820)$. Both of them are in the expected mass range but data for pion emission 
exist for neither.

For negative parity kaons, although we could only compare with two experimental 
ratios, we believe that the evidence supports our assignment of states to multiplets. First, 
our predictions are in good 
agreement with the experimental ratios. Second, one can find reasonable partners for 
$K_3^*(1780)$. Unfortunately, so far, there are no data available for pion emission for 
these partners.
 
\begin{table}
\begin{center}
\begin{tabular}{|c|c|c|} \hline
 Ratio of widths     & Experiment& Heavy quark prediction    \\ \hline
$\frac{\Gamma(K^*(1680)  \rightarrow K +\pi)}{\Gamma( K^*(1680) \rightarrow K^* +\pi)} $ &
$1.3^{+0.23}_{-0.14}$ &1.3\\ \hline\hline
$\frac{\Gamma(K_3^*(1780)  \rightarrow K^* +\pi)}{\Gamma(K_3^*(1780)  \rightarrow K +\pi)} $ &
$1.09\pm0.26$&1.4\\ \hline
\end{tabular}
\caption{Ratio of widths for $K^*(1680)$ in a $(1^-,2^-)$ doublet, and for $K_3^*(1780)$ 
in a $(2^-,3^-)$ doublet. The numbers 
in the second column are the experimental ratios, while the numbers in the third column are the 
HQET predictions.}\label{t3-kaons}
\end{center}
\end{table}

\subsection{Spin-flavor symmetry}\label{fs1}

\subsubsection{Spin-flavor symmetry for the $(1^+,2^+)$ doublets} 

The spin-flavor symmetry tells us that the decay rate of a state from a $(1^+,2^+)$ 
doublet to a state from a $(0^-,1^-)$ doublet is described in terms of a single 
coupling constant, independent of the flavor of the heavy quark. This means that the 
coupling constant describing the processes
\begin{eqnarray}
(K_1,K_2) &\rightarrow&(K,K^*) + \pi,\\
(D_1,D_2) &\rightarrow&(D,D^*) + \pi,\\
(B_1,B_2) &\rightarrow&(B,B^*) + \pi,
\end{eqnarray}
is the same. We can test the extent to which this symmetry holds by examining
the coupling constants for the decays mentioned.

At leading order in HQET, the width of any 
process $A \longrightarrow B+\pi$ is given by
\begin{equation}
\Gamma_i=|\alpha|^2 f_i(m_A,m_B,m_C,\vec{p_\pi})
\end{equation}
where $f_i$ is a known function given by the tensor formalism and phase space. 
Thus, if the spin-flavor symmetry is valid, the ratio 
$\Gamma_i/f_i(m_A,m_B,m_C,\vec{p_\pi})$ should be the same for all the 
processes. Alternatively, we should be able to fit all of these processes using a single value of
$\alpha$.

The results of such a fit are shown in table \ref{fitresults}. In obtaining this fit, we have
averaged the decay rates of the different charge states of the excited $D$ mesons. We note that for
the $D$ mesons, we already mentioned the discrepancy between the ratio of total decay rates of the $D_1$ and
$D_2$ mesons, and the value predicted from HQET. While these results clearly point to the need for
$1/m_Q$ corrections, they are nevertheless somewhat encouraging.

\begin{table}
\begin{center}
\begin{tabular}{|c|c|c|} \hline
 Decay    & Experiment (MeV) & Heavy quark prediction (MeV)   \\ \hline
$K_1(1270)  \rightarrow K^*\pi$ & 14 $\pm$ 6 & 12 \\\hline
$K_2(1430) \rightarrow K\pi$ & 49 $\pm$ 3 & 33 \\\hline
$K_2(1430)  \rightarrow K^*\pi$ & 24 $\pm$ 2 & 33 \\\hline
$D_1  \rightarrow D^*\pi$ & 21 $\pm$ 4 & 13 \\\hline
$D_2\rightarrow D^*\pi+D_2\rightarrow D\pi$ & 24 $\pm$ 4 & 38 \\\hline
\end{tabular}
\caption{Decay widths for states in the $(1^+, 2^+)$ doublet, assuming that the spin-flavor 
symmetry
holds. The second column shows the experimental values, while the third column shows the 
HQET predictions.}
\label{fitresults}
\end{center}
\end{table}
 
 If we use the value of the constant $\alpha$ obtained from this fit, and apply it to the corresponding
 $B$ meson decays, the results we obtain are shown in table \ref{t3-Bmeson}. These results are to
 be compared with those of table~\ref{t2-Bmeson}. All the widths in table~\ref{t3-Bmeson}
 are bigger than the corresponding ones in table~\ref{t2-Bmeson} by about $50\%$. 
 The numbers in the 
 second column are still low, but now the numbers in the fourth column, which were in 
 agreement with experimental data, are too high. The numbers in the third column are 
 now in agreement (within the error bars) with the data.

\begin{table}
\begin{center}
\begin{tabular}{|c|c|c|c|c|} \hline 
 Widths (MeV)    & $m_{B_1}=m_{B_2}=5.68$   &  $m_{B_1}=5.725,m_{B_2}=5.737$  & 
 $m_{B_1}=5.780, m_{B_2}=5.794$ & Experiment\\ \hline 
\(\Gamma( B_2 \rightarrow B +\pi)\)  &$6\pm 1.5$ &$12\pm 3$& $22\pm 8 $&no data\\ \hline
\(\Gamma(B_2 \rightarrow B^* +\pi)\) &$4.5\pm 1.5$ &$10.5\pm 3$&$21\pm 6$ &no data\\ \hline
\(\Gamma( B_2)\)& $10.5\pm 3$ &$22 \pm 6$ &$44 \pm 12$& $25\pm?$\\ \hline
\(\Gamma( B_1 \rightarrow B +\pi) \) &0 &0 &0 & no data \\ \hline
\(\Gamma( B_1 \rightarrow B^* +\pi) \) &$7.5 \pm 3$ & $ 15 \pm 4.5$ & $ 30 \pm 9$ &no data\\ \hline
\(\Gamma( B_1 )\) & $7.5 \pm 3$ & $ 15 \pm 4.5$ & $ 30 \pm 9$ &$20\pm?$\\ \hline
\end{tabular}
\caption{Widths of $B_1$ and $B_2$ mesons using the ``average'' coupling constant obtained from
fitting to the corresponding $D$ and $K$ meson widths.
The numbers of the two first columns are leading order predictions of HQET 
using experimental masses while the numbers of the third column show predictions using masses 
obtained via relations between $B$ mesons and $D$ mesons. The last column shows experimental 
data.}\label{t3-Bmeson}
\end{center}
\end{table}

\subsubsection{Spin-flavor symmetry for $(0^+,1^+)$ doublets} 

After the encouraging results of table~\ref{fitresults} ($i.e.$ applying HQET to the strange 
quark), we can try to glean some information about the widths of the charm and 
the bottom lowest lying $(0^+,1^+)$ doublet using the strange doublet $(0^+, 1^+)$. 
The charm and beauty $(0^+,1^+)$ doublets have not yet been experimentally found, 
mainly because these states are expected to be broad. In what follows, we use masses
obtained using a quark model \cite{ZVD}. The quark model masses for these states are 
respectively $m_{D_0}=2270$ MeV, $m_{D_1}=2400$ MeV, $m_{B_0}=5.65$ GeV and 
$m_{B_1}=5.69$~GeV.

Using experimental decay rates associated with the strange $(0^+,1^+)$ doublet 
$(K_0^*(1430)$, $K_1(1400))$, we can predict the widths for the corresponding charm and 
beauty $(0^+,1^+)$ doublets. These are shown in table~\ref{t1-fs}.

The HQET predictions, decay rates around $70$ MeV, are somewhat smaller than one expects from 
most quark models. The model of Goity and Roberts \cite{WinstonQM} predicts widths of
about 120 MeV, while most other models predict much larger widths \cite{gk}. 
The differences between the two sets of
quark model predictions have been attributed to relativistic effects \cite{WinstonQM}. What is
significant, we believe, is that the HQET predictions are of similar size to those predicted by
reference \cite{WinstonQM}. Nevertheless, there are a number of possibilities for generating
larger widths using HQET.

The HQET predictions are quite sensitive to the masses, and if we increase the 
masses of $D_0, D_1$ and $B_0, B_1$ by $100$ MeV one finds widths of the order of $120$ 
MeV for $D_0$ and $D_1$ and 140 MeV for $B_0$ and $B_1$, very much in agreement with the work 
of Goity and Roberts \cite{WinstonQM}. Perhaps here we have a hint that the quark model predicted 
masses of $D_0, D_1$ and $B_0, B_1$ are too low. Another possibility is that our 
identification of $(K_0^*(1430), K_1(1400))$ as the lowest lying $(0^+,1^+)$ doublet is incorrect. 
This would make it meaningless to extrapolate the coupling constant to the corresponding 
charm and beauty mesons. Finally, one can assume that the doublet identification is 
correct but that $1/m_Q$ corrections are very important (especially for the strange 
$(0^+,1^+)$ doublet 
$(K_0^*(1430), K_1(1400))$, and that if we included these corrections, we could predict larger 
widths. 

\begin{table}[h]
\begin{center}
\begin{tabular}{|c|c|c|} \hline
 Widths (MeV) &$m_{D_0}=2.27,m_{D_1}=2.40$ GeV &$m_{D_0}=2.37,m_{D_1}=2.50$ GeV\\ \hline
$\Gamma(D_0  \rightarrow D +\pi)$  & $70 \pm 5$ & $120 \pm 10$ \\ \hline
$\Gamma(D_1  \rightarrow D^*+\pi)$ & $67 \pm 5$ & $120 \pm 10$ \\ \hline\hline
   &$m_{B_0}=5.65,m_{B_1}=5.69$ GeV &$m_{B_0}=5.75,m_{B_1}=5.79$ GeV\\ \hline
$\Gamma(B_0  \rightarrow B +\pi)$  & $70 \pm 5$ & $140 \pm 12$ \\ \hline
$\Gamma(B_1  \rightarrow B^*+\pi)$ & $67 \pm 5$ & $140 \pm 12$ \\ \hline
\end{tabular}
\caption{Predictions of the decay rates of the lowest lying 
 $(0^+,1^+)$ charm and bottom doublets using spin-flavor symmetry. }\label{t1-fs}
\end{center}
\end{table}

\section{Conclusion} \label{sec4}

In the previous sections, we have used the heavy quark tensor formalism to analyze 
strong decays of excited heavy hadrons. We have compared experimental and theoretical 
ratios of decay rates for $B$ meson, $D$ meson and kaons. We have not compared our
predictions with data in the baryon sector, as baryon data are either not yet
sufficiently precise, or not yet measured, for such comparison to be meaningfully made. 
Our results 
are in agreement with the spin-counting methods of Isgur and Wise. We have found some 
encouraging results if we treat the strange quark as heavy. Nevertheless, in this case, 
one would certainly expect $1/m_Q$ corrections to be very important. For such reasons 
we believe that terms of subleading order in the heavy quark expansion should be studied.
 In addition, the experimental situation in all sectors needs to be improved before we 
 can make more precise tests of the predictions of HQET. In the strange sector, 
 experiments that 
 will be carried out at the Brookhaven National Laboratory, and perhaps also at Jefferson
 Laboratory, should help remedy this situation.
 
\section*{Acknowledgement}

The support of the National Science Foundation through grant 947582, and of the
Department of Energy through grants DE-FG05-94ER40832 and DE-AC05-84ER40150 is
gratefully acknowledged.

\end{document}